\documentclass[amsmath,amssymb,aip,apl,twocolumn,11pt,reprint,superscriptaddress]{revtex4-1}
\usepackage{graphicx}
\usepackage{pstricks}
\usepackage{epstopdf}

\DeclareMathOperator{\cn}{cn}

\DeclareMathOperator{\dn}{dn}

\DeclareMathOperator{\sn}{sn}
\DeclareMathOperator{\di}{d\kern-0.4ex}

\begin{document}
%\preprint{}

\title{Symmetric metal slot waveguides with nonlinear dielectric core: bifurcations, size effects, and higher order modes}

\author{Wiktor Walasik}
\affiliation{Aix---Marseille Universit\'{e}, CNRS, Ecole Centrale, Institut Fresnel, 13013 Marseille France}
\email{gilles.renversez@fresnel.fr}
\homepage[]{www.fresnel.fr/spip/clarte}
\affiliation{ICFO --- Institut de Ci\`{e}ncies Fot\`{o}niques, Universitat Polit\`{e}cnica de Catalunya, 08860 Castelldefels (Barcelona), Spain}
\author{Alejandro Rodriguez}
\affiliation{Departement of Electrical Engineering, Princeton University, NJ 08544, U.S.A.}
\author{Gilles Renversez}
\affiliation{Aix---Marseille Universit\'{e}, CNRS, Ecole Centrale, Institut Fresnel, 13013 Marseille France}
\email{gilles.renversez@fresnel.fr}
\homepage[]{www.fresnel.fr/spip/clarte}
%altaffiliation{1}

\begin{abstract}
We study the nonlinear waves propagating in metal slot waveguides with a Kerr-type dielectric core. We develop two independent semi-analytical models to describe the properties of such waveguides. Using those models we compute the dispersion curves for the first ten modes of a nonlinear slot waveguide. 
For symmetric waveguides we find symmetric, antisymmetric, and asymmetric modes which are grouped in two families. In addition, we study the influence of the slot width on the first symmetric and asymmetric modes, and we show that the dispersion curve of the first asymmetric mode is invariant with respect to the slot  width for high propagation constant values and we provide analytical approximations of this curve.  
\end{abstract}

\date{\today}
%\pacs{42.65.Wi, 42.65.Tg, 42.65.Hw, 73.20.Mf}
%\keywords{Nonlinear waveguides, optical, Optical solitons, Kerr effect: nonlinear optics, Plasmons on surfaces and interfaces / surface plasmons}

\maketitle

Nonlinear plasmonics is now a flourishing branch of photonics~\cite{Kauranen12}. Among it,  the studies of nonlinear waves combining both plasmon and soliton features started in 1980s~\cite{Ariyasu85,Mihalache89}. The interest in the field started to grow again thanks to the results provided in Ref.~\cite{Feigenbaum07}. Even if no experimental demonstration of plasmon--solitons has been published yet~\cite{Walasik12}, in the mid term, such states can have several applications, e.g.\ in nonlinear couplers~\cite{Salgueiro10} or in a four wave mixing process~\cite{Renger09}.
Moreover, linear metal slot waveguides with a few hundred nanometers wide silicon core have already been fabricated and characterized~\cite{Han:10,1.4772941}. 

The structure studied in Ref.~\cite{Feigenbaum07} is a nonlinear dielectric core surrounded by two semi-infinite metal regions. It will be called here the nonlinear slot waveguide (NSW).
Recent studies of plasmon--solitons in NSWs with Kerr-type focusing nonlinear core have led to a number of interesting predictions, including the existence of symmetric, antisymmetric, and asymmetric modes~\cite{Davoyan08,Rukhlenko11a}. 
The asymmetric nonlinear mode appears from the symmetric one through a symmetry-breaking bifurcation at a critical power. This bifurcation phenomenon has already been described in nonlinear slot waveguides, fully dielectric nonlinear waveguides, and waveguides made of nonlinear metal sandwiched between linear dielectrics~\cite{Boardman86a,Holland86,Kivshar01,Davoyan08,Davoyan11}.

 In this letter, we demonstrate the existence of a number of previously unknown higher order modes in the NSW. We show that bifurcations also occur for some of the higher order modes. We study the influence of the slot width on the dispersion curves of the first asymmetric mode. We show that these curves have an invariant part that can be described analytically. 

These results are obtained using two approaches we developed: (i) a semi-analytical model, where field profiles in the core are described by the Jacobi elliptic functions and (ii) a model where the field profile in the core is obtained by direct integration of Maxwell's equations.

\begin{figure}[htbp]
\centerline{\includegraphics[width = 0.8\columnwidth,angle=-0,clip=true,trim= 0 0 0 0]{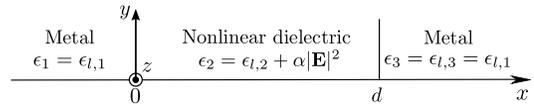}}
\caption{Symmetric nonlinear slot waveguide geometry.}
\label{fig:geom}
\end{figure}

Let us consider monochromatic transverse magnetic (TM) waves propagating along the $z$ direction (all field components evolve proportionally to ($\exp[i( k_0 \beta z - \omega t)]$) in a symmetric one-dimensional NSW depicted in Fig.~\ref{fig:geom}. Here $k_0 = \omega/c$, where $c$ denotes the speed of light in vacuum, $\beta$ denotes the effective mode index and $\omega$ is the light angular frequency. 

The nonlinear Kerr-type dielectric is isotropic with 
$\epsilon = \epsilon_l + \alpha |\textbf{E}|^2$ and $\alpha > 0$ (focusing nonlinearity). The relation between the nonlinear parameter $\alpha$ and the coefficient $n_2$, that appears in the definition of an intensity dependent refractive index $n = \sqrt{\epsilon_{l}} + n_2 I$, is $\alpha  = \epsilon_0 c \epsilon_{l} n_2$
 where the intensity is defined as $I = \epsilon_0 c |\epsilon| |\textbf{E}|^2 /(2 \beta)$, and the vacuum permittivity is denoted by $\epsilon_0$.

Our first model uses a semi-analytical approach that provides closed analytical expressions for the nonlinear dispersion relation and for the field profiles of the modes. 
It employs the fact that Maxwell's equations for TM electromagnetic wave in the Kerr-type nonlinear medium can be reduced to a single nonlinear wave equation for the magnetic field component~\cite{Ariyasu85,Walasik14}, assuming that the nonlinear permittivity modification is small and depends only on the transverse $E_x$ component. This equation is solved using the first integral approach~\cite{Chen88,Fedyanin82} yielding  
\begin{equation}
\left(\frac{d H_y}{d x}\right)^2 - k_0^2 q_2^2 H_y^2 + k_0^2 \frac{a}{2} H_y^4 = c_0,
\label{eqn:wave}
\end{equation} 
where $q_i^2 = \beta^2-\epsilon_{l,i}$ ($i \in \{1,2,3\}$ enumerates the region), $a =\beta^2 n_2 /(\epsilon_0 \epsilon_{l,2} c)$, and $c_0$ is the integration constant that in the case of a finite-size nonlinear medium is given~by
\begin{equation}
c_0 = k_0^2 \big[(  {\epsilon_{2}}/{\epsilon_1})^2 q_1^2 - q_2^2 + {aH_y(0)^2}/{2}     \big] H_y(0)^2.
\label{eqn:c0}
\end{equation} 
The resulting field profiles in the NSW core are described by Jacobi elliptic special functions~\cite{Abramowitz72} and hence we denote this model as the Jacobi elliptic model (JEM).

The problem is split in two cases depending on the sign of $c_0$. For $c_0\le0$ only node-less solutions exist. In contrast, for $c_0>0$ only solutions with nodes appear.
The field profile for the case $c_0>0$ is given by:
\begin{equation}
H_y(x) = \delta_+  \cn\kern-0.6ex\big[\sqrt{{{s}/{A}}}\; (x-x_0)    | m \big],
\end{equation}
in which
\begin{align}
\delta_\pm^2 &= \big(\sqrt{A^2Q^2 + 4 A c_0} \pm AQ\big)/2, \\
x_0 &= - \sqrt{{A}/{s}}   \cn^{-1}\kern-0.6ex\big[{H_y(0)}/{\delta_+} | m\big]. \label{eqn:x0}
\end{align}
Here $Q = k_0^2 q_2^2$, $A = 2/(k_0^2 a)$,  $s = \delta_+^2 + \delta_-^2$, $m = {\delta_+^2}/{s}$, and $\cn[x|m]$ a Jacobi elliptic function with  a parameter~$m$~\cite{Abramowitz72}. 

Using the continuity of the tangential field components at the interfaces between nonlinear medium and the bounding metals, were fields decay exponentially, the nonlinear dispersion relation is obtained in an analytical form (for $c_0>0$):
\begin{align}
{k_0 q_3 \epsilon_{2}} \sqrt{{A}/{s}}
\cn\kern-0.6ex\big[ \sqrt{{{s}/{A}}}\; (d-x_0)    | m \big] = \;\;\;\;\;\;\;\;\;\;\;\;\;\;\;    \nonumber \\
\epsilon_3\sn\kern-0.6ex\big[  \sqrt{{{s}/{A}}}\; (d-x_0)    | m \big]
\dn\kern-0.6ex\big[  \sqrt{{{s}/{A}}}\; (d-x_0)    | m \big],
\label{eq:disp-JEM}
\end{align} 
in which Jacobi elliptic functions $\sn$ and $\dn$ appear. For a given structure the solutions of Eq.~(\ref{eq:disp-JEM}) are found fixing the values of the free parameter $H_y(0)$ [see Eq.~(\ref{eqn:x0})].

Similar equations can be derived for the case $c_0<0$ \linebreak but the field profile and dispersion relations are described by different Jacobi elliptic functions.

We also develop a second model, based on the results of Refs~\cite{Mihalache87, Yin09} for a single nonlinear dielectric/metal interface only, where the nonlinear term accounts for both transverse and longitudinal electric field components. This approach allows one to obtain a closed-form analytical expression for the dispersion relations for each of the two interfaces of the NSW. In particular, for the left interface, one obtains:
\begin{align}
\Big\{ \big[ (  {\epsilon_{2,0}}/{\beta}  )^2 - 2\epsilon_{2,0}   \big]& {(\epsilon_1 \beta)^2}/{\big[(\epsilon_1 \beta)^2 +(\epsilon_{2,0} q_1)^2\big]} \nonumber \\ + &\epsilon_{l,2} + {\epsilon_0 c \epsilon_{l,2} n_2 E_0^2}/{2}  \Big\} E_0^2 = C,
\label{eqn:EM}
\end{align}
where
$E_0 = \sqrt{E_{x}^2(0) + E_{z}^2(0)}$ denotes the total electric field at the left interface and $\epsilon_{2,0} = \epsilon_{l,2} + \alpha E_0^2$. To get the equation on the right interface (see Fig.~\ref{fig:geom}) we substitute $E_0$ by $E_d$ and $\epsilon_{2,0}$  by $\epsilon_{2,d}$.

For fixed  $\lambda$, $d$, and material parameters ($\epsilon_{l,1}, \epsilon_{l,2}, n_2$) the dispersion relation is computed using a shooting method. Generally, the three dimensional phase space of possible solutions spanned by $E_0$, $E_d$ and $\beta$ is scanned. For each triplet a set of Maxwell's equations coupling $E_x$ and $E_z$ components~\cite{Mihalache87,Yin09,Walasik14} 
\begin{subequations}
\label{eqn:yin_system}
 \begin{align}
\frac{d E_z}{d x} &= k_0\left( \beta - \frac{\epsilon_2}{\beta} \right)E_x,
\label{eqn:yin1}\\
 \frac{d E_x}{d x} &= k_0 \frac{\beta  \epsilon_2  E_z - 2  \alpha_1  E_z  E_x^2  \left(\beta - \frac{\epsilon_2}{\beta}  \right)}{ \epsilon_2 + 2  \alpha  E_x^2}
\label{eqn:yin_fields_system2}
\end{align} 
\end{subequations}
is solved by numerical integration inside the waveguide core. If the solution in the core is consistent with the previously assumed field values at the slot interfaces then the corresponding $\beta$ is accepted as a genuine solution of the problem. This model will be called the interface model (IM).

In our case of a symmetric slot waveguide the problem can be split in two cases: (i) symmetric and antisymmetric modes and (ii) asymmetric modes. In the first case the phase space can be reduced to two dimensions ($E_0$ and $\beta$) using the symmetry argument ($E_d = \pm E_0$). In the second case the phase space can also be reduced to two dimensions (this time $E_0$ and $E_d$) using the following method.
Equating right-hand sides of the dispersion equation [Eq.~(\ref{eqn:EM})] for the left interface with the one for the right interface, allows one to eliminate the integration constant $C$. It follows that for the fixed field intensities at both interfaces ($E_0$ and $E_d$), this equation can be solved analytically providing the allowed values of the propagation constant $\beta$. The resulting equation for $\beta$ has the form $a_4 \beta^4 + a_2 \beta^2 + a_0 = 0$ which means that we can have up to two different possible solutions for a given pair ($E_0$, $E_d$) ($\beta$ being real and positive) that have to be verified using the numerical integration of Eqs.~(\ref{eqn:yin_system}).

\begin{figure}[!ht]
\centerline{\includegraphics[width=0.95\columnwidth,clip=true,trim= 3 2 10 2]{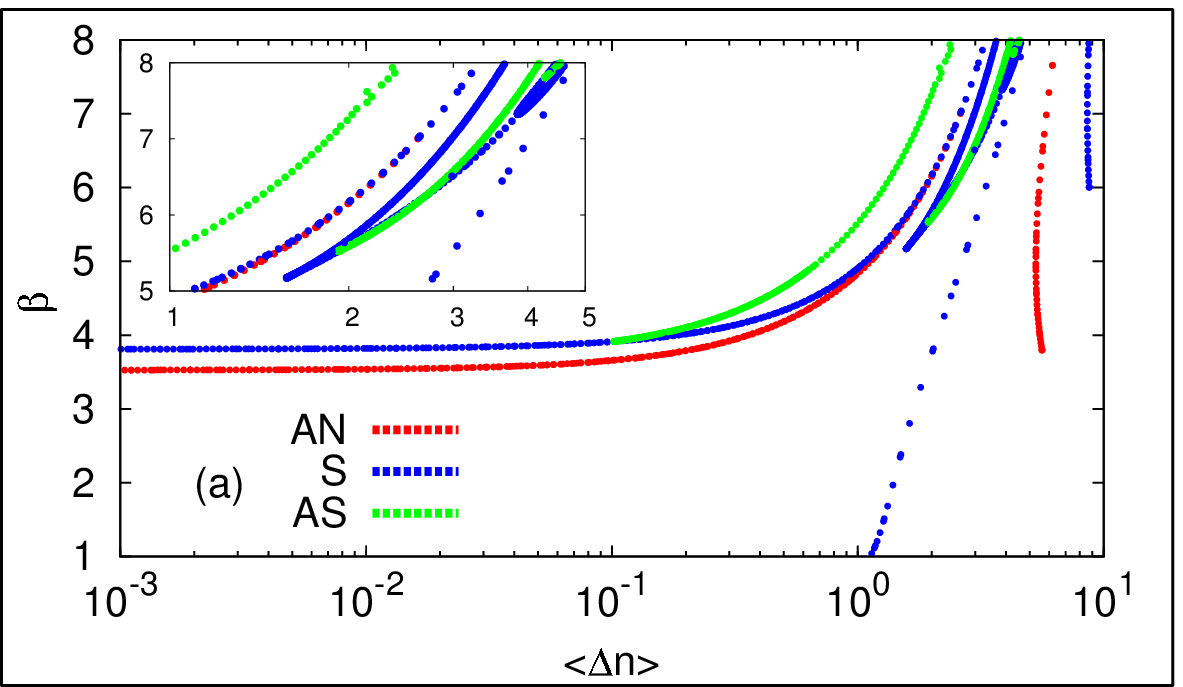}}
\centerline{\includegraphics[width=0.95\columnwidth,clip=true,trim= 3 2 10 2]{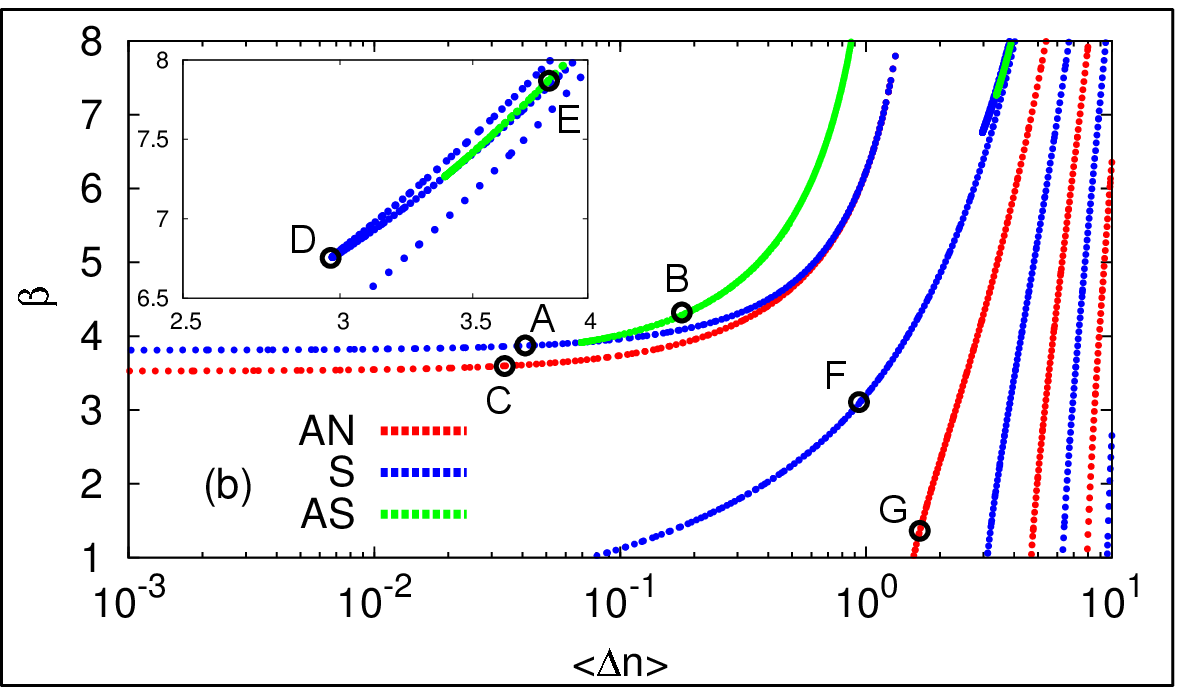}}
\caption{Dispersion curves of the first and higher order symmetric (S --- blue), antisymmetric (AN --- red), and asymmetric (AS --- green) modes for the symmetric NSW obtained with the JEM (a) and the IM (b). The insets show zooms on the regions with bifurcation of the first symmetric higher order mode. Results obtained for a $0.4$-$\mu$m-wide hydrogenated amorphous silicon core $\epsilon_{l,2} = 3.46^2$ ($n_2 = 2\cdot 10^{-17}$ m$^2$/W) \cite{Matres12} sandwiched between gold $\epsilon_1 = \epsilon_3 = -90$ at a wavelength $\lambda = 1.55$~$\mu$m. Losses in metals are neglected.}
\label{fig:disp}
\end{figure}

Figure~\ref{fig:disp} shows the NSW dispersion curves for solutions corresponding to different symmetries, obtained by  (a) the JEM and (b) the IM, as a function of the averaged nonlinear index modification $\langle\Delta n\rangle = 1/d \int^d_0 n_2 I \di x$.  In the region where $\langle\Delta n\rangle<0.1$  we find good quantitative agreement between results of these two models, which are radically different in this respective approach to deal with NSWs.  
In this region the electric field component peak ratio $\max\{|E^2_z|\}/\max\{|E^2_x|\}$ is smaller than $0.25$. For values of $\langle\Delta n\rangle$ above $0.1$ both models predict qualitatively the same type of behavior but lead to quantitative differences, that can be understood looking at the assumptions used in the JEM. Such results confirm the validity of both models and of their numerical implementation. It is worth mentioning that even if the IM offers more generality than the JEM, the latter requires ten times less computational time than the former one. The IM also provides more physical insights since the field profiles are described analytically even if it employs unusual special functions.   
In what follows, we describe results obtained with the IM.

\begin{figure}[!ht]
\begin{center}
\includegraphics[width=0.49\columnwidth,clip=true,trim= 0 0 0 0]{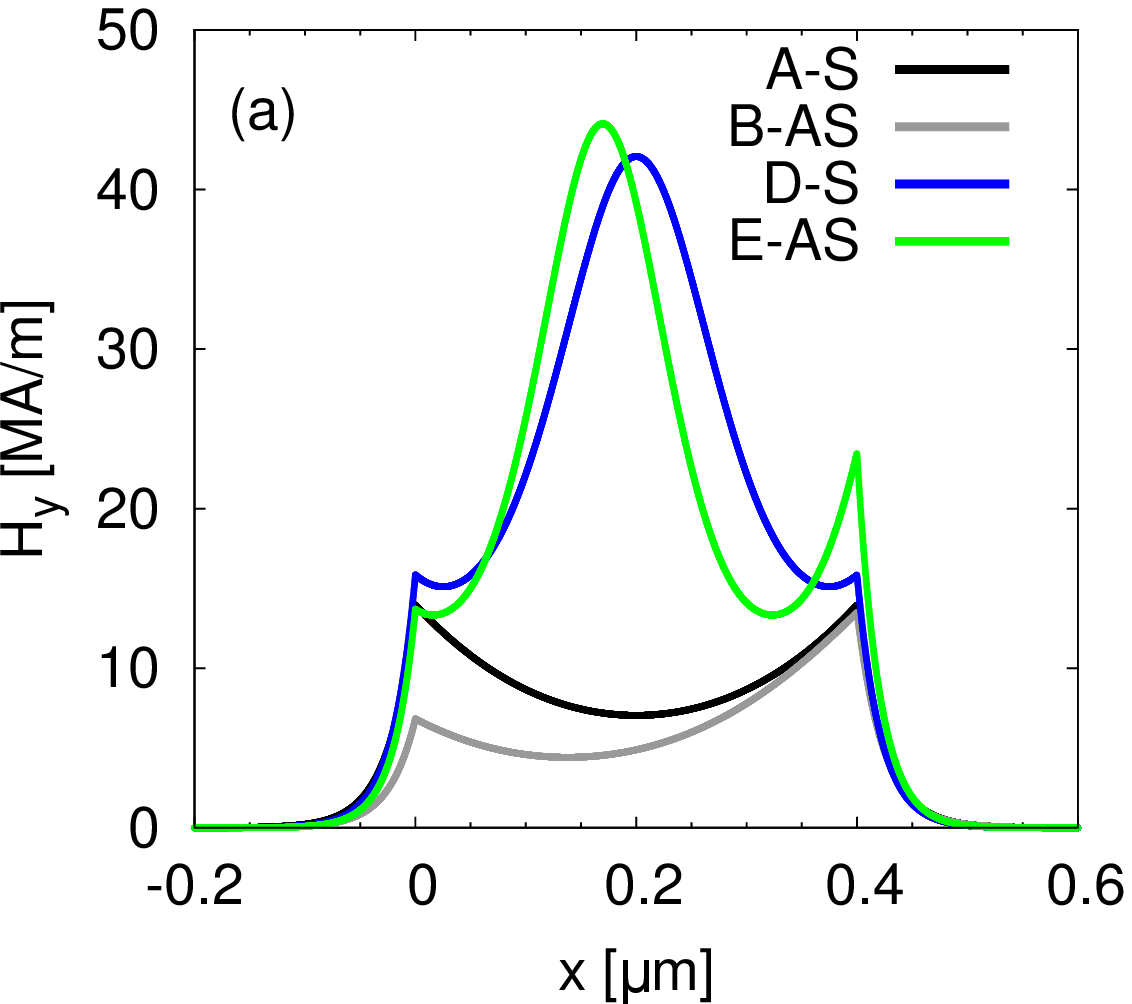}
\includegraphics[width=0.49\columnwidth,clip=true,trim= 0 0 0 0]{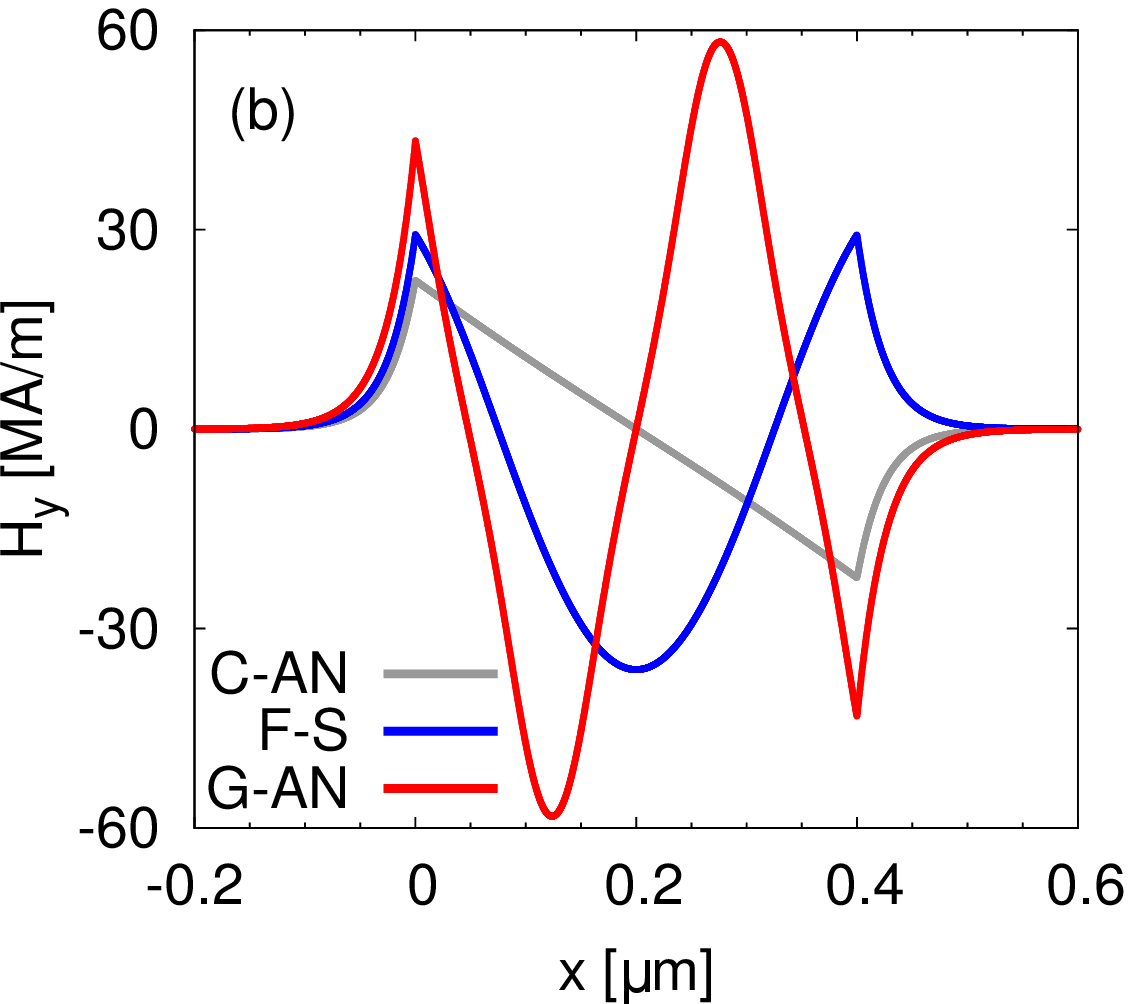}
\caption{Magnetic field component for the first and higher order node-less modes (a) and the modes with nodes (b)  corresponding to the points indicated in Fig.~\ref{fig:disp}(b).}
\label{fig:fields}
\end{center}
\end{figure}

For low light intensity, i.e.\ small index modification, a symmetric and an antisymmetric mode are obtained \cite{Rukhlenko11a}.
At $\langle\Delta n\rangle \approx 0.07$ an asymmetric mode bifurcates from the symmetric one~\cite{Davoyan08}. The effective index $\beta$ of the three modes described here grows rapidly with the increase of $\langle\Delta n\rangle$. 
The field profiles of these first order modes are presented in gray and black in Fig.~\ref{fig:fields}. For larger $\langle\Delta n\rangle$ two groups of higher order modes appear. The first group of node-less modes, appearing for $\langle\Delta n\rangle \gtrsim  3$ in the frame of IM, possesses only very high effective indices $\beta$.
The insets in Fig.~\ref{fig:disp} show enlarged regions of parameter space where these modes appear, and where bifurcation of asymmetric mode from symmetric higher-order mode occurs. 
The typical field profiles of these higher order modes are shown in Fig.~\ref{fig:fields}(a). Both the first and the second asymmetric branch are doubly degenerated (i.e. at fixed $\beta$ there exist two modes with equal powers but shapes inverted with respect to the center axis $x=d/2$) but for simplicity we show only one of the field plots. The second group of higher order modes with nodes, appearing for $\langle\Delta n\rangle \gtrsim  0.08$ in the frame of IM,  consists of an alternation of symmetric and anti-symmetric modes. These modes resemble the higher order modes of a linear slot waveguide with $\epsilon_{l,2}>3.46^2$. The field profiles of the two first modes of this second group are provided in Fig.~\ref{fig:fields}(b).

\begin{figure}[!b]
\centerline{\includegraphics[width=0.95\columnwidth,clip=true,trim= 0 0 0 0]{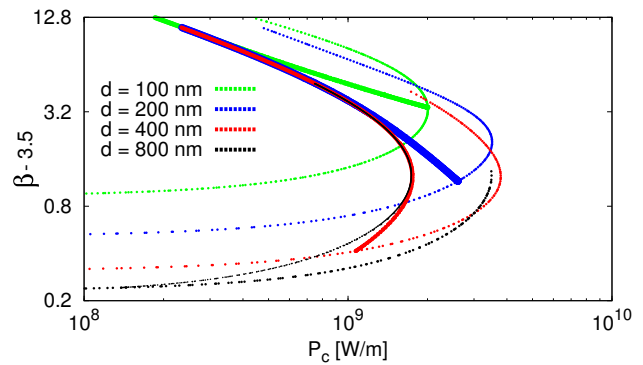}}
\caption{Locus of the asymmetric mode dispersion curves computed with the IM for various core widths $d$ as a function of the power localized in the core $P_c$. Only first symmetric (thin curve) and asymmetric modes (thick curves, except for the black one) are shown using one color per width. Other parameters are the same as in Fig.~\ref{fig:disp}. Both axes are in log scale.}
\label{fig:locus}
\end{figure}

Figure~\ref{fig:locus} shows the dispersion curves for the parameter space region where asymmetric mode bifurcates from symmetric mode  for different widths of the NSW core. 
Notice that the effective mode index is plotted as a function of the power localized in the core $P_c= \int^{d}_{0} \mathcal{S}_z \di x$, in which $ \mathcal{S}_z$ denotes the $z$-component of the Poynting vector. 
One can observe that, in this coordinate, the $\beta(P_c)$ curves for asymmetric modes have invariant parts with respect to the core width. The same conclusion is also obtained if one considers the intensity density $I_d = \int^{+\infty}_{-\infty} I \di x$ as abscissa (data not shown). The overlap of the  nonlinear dispersion curves is obtained for large differences of the effective indices between symmetric and asymmetric modes $\Delta \beta$.

This is the consequence of the fact that with the increase of effective index the asymmetric modes concentrate around only one of the interfaces, so that the problem reduces to a single metal/dielectric interface where the width of the core is no longer important. 

As it can be expected the curves corresponding to zero integration constants $c_0=0$ in the JEM [see Eq.~(\ref{eqn:c0})]  provides solutions of the  single metal/nonlinear dielectric interface problem.
Using Eq.~(\ref{eqn:c0}) of the JEM, an analytical formula for the approximated effective index of the highly asymmetric modes  is derived for  large $\Delta \beta$ values 
\begin{equation}
\widetilde{\beta} = \sqrt{{\epsilon_1\epsilon_{l,2}(\epsilon_{l,2}-\epsilon_1)}/\big[{\epsilon_{l,2}^2-\epsilon_1^2 + {n_2\epsilon_1^2 H_0^2}/{(2 \epsilon_0 c \epsilon_{l,2})}\big]}},
\label{eqn:beta_approx}
\end{equation} 
where $\epsilon_2$ was approximated by $\epsilon_{l,2}$.

\begin{figure}[!t]
\begin{center}
\includegraphics[width=0.49\columnwidth,clip=true,trim= 0 0 0 0]{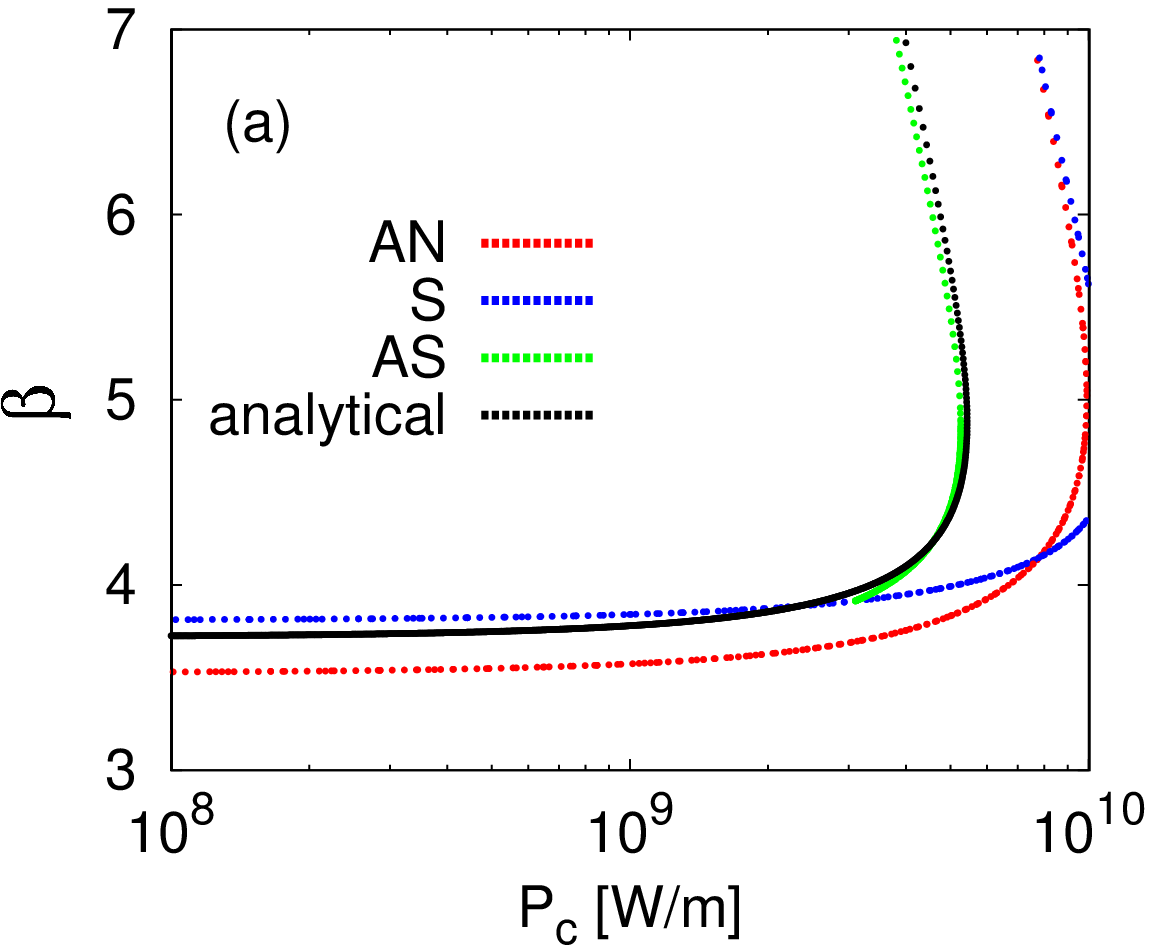}
\includegraphics[width=0.49\columnwidth,clip=true,trim= 0 0 0 0]{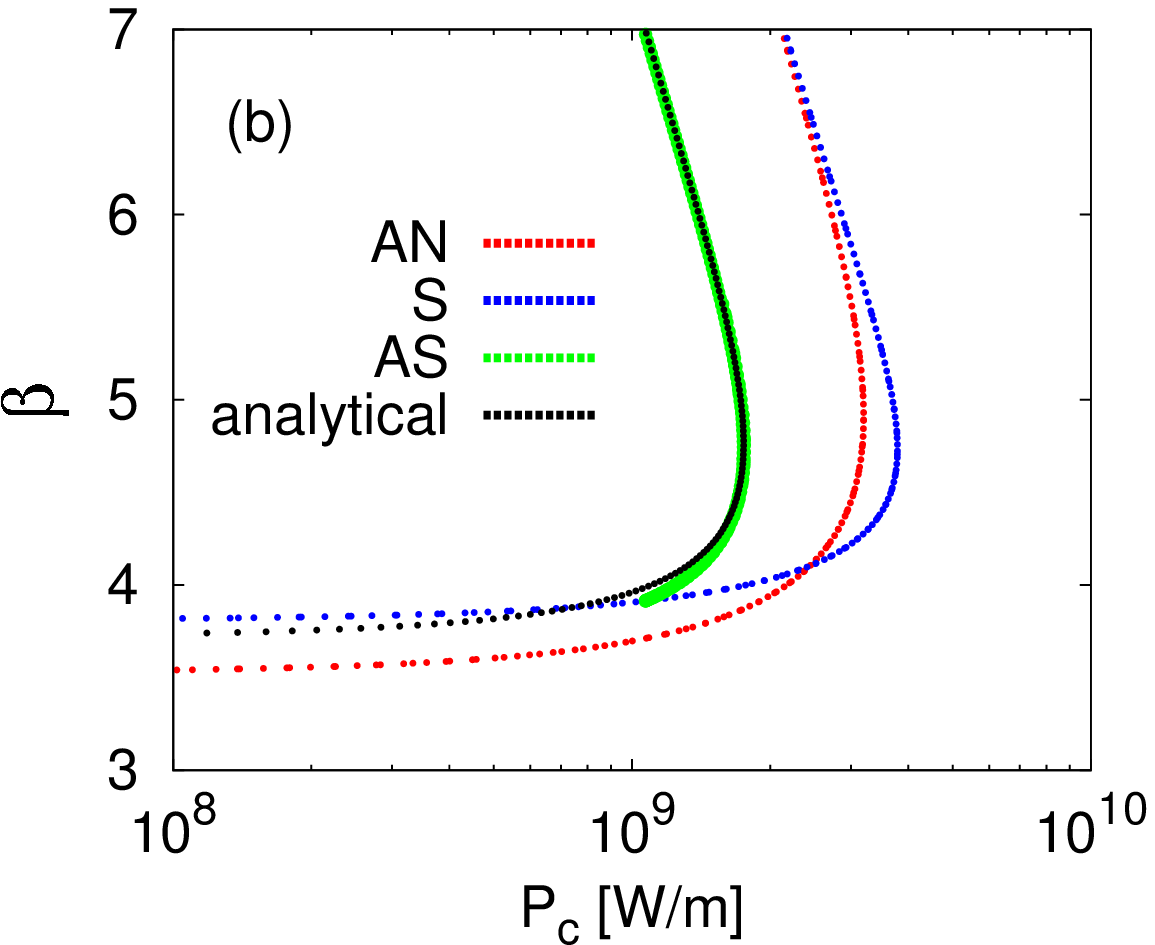}
\caption{Zoom of the dispersion curves of the first  symmetric (S --- blue), antisymmetric (AN --- red), and asymmetric (AS --- green) modes for the symmetric NSW obtained with the JEM (a) and the IM (b). In each case, the analytical approximation corresponding to a single interface problem is shown with a black dotted line. All the parameters are the same as in Fig.~\ref{fig:disp}.}
\label{fig:fit}
\end{center}
\end{figure}

For the IM, to describe the single metal/nonlinear dielectric interface problem, one has to set $C=0$ in Eq.~(\ref{eqn:EM}).
The analytical solution of this equation is provided by Eq.~(14) in Ref.~\cite{Yin09}. As it can be seen in Fig.~\ref{fig:fit}, both approximations nicely fit their respective computed dispersion curves of asymmetric modes.

\begin{figure}[!t]
\centerline{\includegraphics[width=0.75\columnwidth,clip=true,trim= 0 0 0 0]{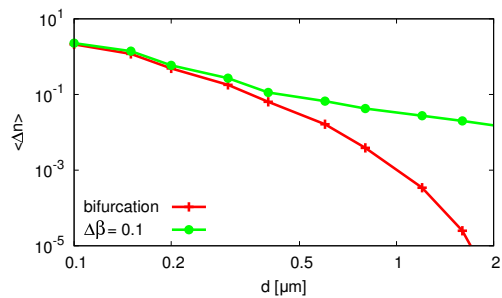}}
\caption{Average nonlinear index change $\langle\Delta n\rangle$ at the appearance of the asymmetric modes (bifurcation) (red~$+$) and for $\Delta \beta = 0.1$ (green~$\bullet$) as a function of the core width $d$. }
\label{fig:bif}
\end{figure}

Figure~\ref{fig:bif} shows the dependence of the nonlinear index change $\langle\Delta n\rangle$ at which the first bifurcation occurs (red curve) on the width of the NSW core. One can see that the thicker the core the lower the value of $\langle\Delta n\rangle$ where the bifurcation occurs. For $d\approx 1$~$\mu$m (being $\approx 2 \lambda/3$) the bifurcation occurs at $\langle\Delta n\rangle$ between $10^{-5}$ and $10^{-3}$ which are realistic values for hydrogenated amorphous silicon~\cite{Matres12}. The value of $\langle\Delta n\rangle$ for which $\Delta \beta = 0.1$ is shown in green. The values of $\langle\Delta n\rangle$ for which the asymmetric modes become strongly asymmetric, i.e.\ light intensity of the asymmetric mode at one metal/dielectric interface is five times larger than on the other one, is approximately twice higher than the bifurcation threshold. Cladding with higher permittivity also allows to reduce the $\langle\Delta n\rangle$ where the bifurcation occurs, e.g.\ for $0.4$-$\mu$m-wide core sandwiched between metals with permittivity equal to $-40$ (all other parameters unchanged) $\langle\Delta n\rangle$ is reduced 4 times and for metals with permittivity equal to $-20$ $\langle\Delta n\rangle$ is reduced 100 times.

In conclusions, we have developed two independent models describing nonlinear slot waveguides. Qualitatively the results of the simplified, but analytical approach and the exact more numerical approach are the same. Both models predict bifurcation of asymmetric modes from the symmetric node-less modes. We show that the bifurcation occurs not only for the first but also for higher order nonlinear modes. Both the nonlinear dispersion curves and the field profiles for the higher order symmetric and antisymmetric modes are shown for the first time. We show that the first nonlinear asymmetric mode dispersion curves have invariant parts with respect to the core width and we provide analytical formula for this invariant curve.
We also obtain that the intensity at which the bifurcation occurs decrease with the increase of the core width or the metal permittivity. 

%\begin{acknowledgments}
This work was supported by the European Commission through the Erasmus Mundus Joint Doctorate Programme  Europhotonics (Grant No.\ 159224-1-2009-1-FR-ERA MUNDUS-EMJD). The authors thank Y.~V.~Kartashov for helpful comments on this work. G.~R.~ thanks J.-M.~F\'{e}d\'{e}li for fruitful discussions.
%\end{acknowledgments}

%\begin{acknowledgments}
% Put your acknowledgments here.
%\end{acknowledgments}

%%%%%%%%%%%%%%%%%%%%%%% References %%%%%%%%%%%%%%%%%%%%%%%%%
%
\end{document}